%% file: Poster__testing_network_based_RTK_for_GNSS_robustenss_and_security.tex
\documentclass[sigconf,screen,nonacm]{acmart}

\acmConference[WiSec '24]{WiSec '24: 17th ACM Conference on Security and Privacy in Wireless and Mobile Networks}{May 27--May 30, 2024}{Seoul, Korea}
\acmBooktitle{WiSec '24: 17th ACM Conference on Security and Privacy in Wireless and Mobile Networks, May 27--May 30, 2024, Seoul, Korea}
\acmPrice{15.00}
\acmISBN{978-1-4503-XXXX-X/21/06}

\usepackage{glossaries}
\usepackage{siunitx}
\usepackage{caption}
\usepackage{subcaption}
\usepackage[capitalise]{cleveref}
\glsdisablehyper %% http://tug.ctan.org/macros/latex/contrib/glossaries/glossariesbegin.pdf -> disable highlighted gls entries in screen version, clickable ones for all versions
	
\input{acronyms.tex}

\begin{document}

\title{POSTER: Testing network-based RTK for GNSS receiver security}
% if too long use  \title[short title]{full title}
%\subtitle{how to spoof gnss signals via replay}

\author{Marco Spanghero}
\orcid{0000-0001-8919-0098}
\affiliation{%
	\institution{Networked Systems Security Group \\ KTH Royal Institute of Technology}
	\streetaddress{Isafjordsgatan 26, 16440 Kista}
	\city{Stockholm}
	\country{Sweden}
}
\email{marcosp@kth.se}

\author{Panagiotis Papadimitratos}
\orcid{0000-0002-3267-5374}
\affiliation{%
	\institution{Networked Systems Security Group \\ KTH Royal Institute of Technology}
	\streetaddress{Isafjordsgatan 26, 16440 Kista}
	\city{Stockholm}
	\country{Sweden}
}
\email{papadim@kth.se}

%\renewcommand{\shortauthors}{Lenhart et al.}

\begin{abstract}
Global Navigation Satellite Systems (GNSS) provide precise location, while \gls{rtk} allow mobile receivers (termed rovers), leveraging fixed stations, to correct errors in their \gls{pnt} solution. This allows compensating for multi-path effects, ionospheric errors, and observation biases, enabling consumer receivers to achieve centimeter-level accuracy. While network distribution of correction streams can be protected with common secure networking practices, the reference stations can still be attacked by GNSS spoofing or jamming. This work investigates (i) the effect \gls{rtk} reference station spoofing has on the rover's \gls{pnt} solution quality and (ii) the potential countermeasures towards hardening the \gls{rtk} infrastructure.
\end{abstract}

\maketitle

\section{Introduction}
\gls{gnss} are ubiquitous and provide localization and timing for a wide gamut of often strategic location-based services. For precision navigation, \gls{rtk} leverages multiple receivers to correct \gls{gnss} measurements at the mobile station (rover) using differential ranging and a known baseline with a reference station (base). Specifically, network-based \gls{rtk} is established as an open-source, collaborative system to achieve centimeter-level accuracy with consumer-grade receivers.

Nevertheless, the current unencrypted and public nature of the \gls{gnss} signals make \gls{gnss} receivers a relatively easy target for manipulation, via spoofing, meaconing (e.g. replay/relay), or jamming \cite{humphreysAssessingSpoofingThreat2008, LenhartSP:C:2022, Motallebighomi2023}. Although cryptographic countermeasures to the spoofing problem exist, adopting such methods will take time especially when they require modifications to the signal in space or the receiver structure \cite{papadimitratosProtectionFundamentalVulnerability2008}. 

Generally, a rover connects to the closest reference station to the area it operates in, and the GNSS corrections are considered meaningful within a \SI{10}{\kilo\meter} radius of the reference station.
While the network-based correction stream can effectively be protected using secure internet protocols, to avoid manipulation of the information during transfer, e.g., see \cite{Pepjin:2020}, these methods fall short if the adversary can directly manipulate the \gls{gnss} signals at the reference station. Even low-sophistication spoofing attacks are possible with low-cost hardware and open-source implementation, in particular, if simulation or replay/relay-based and can effectively control a \gls{gnss} receiver \cite{gpspatron}.

In this work, we evaluate the effects of different types of interference, from simplistic barrage jamming to sophisticated synchronous lift-off on an \gls{rtk} base station, and how such manipulation reflects on the victim receiver. We attack our reference station, to avoid causing disturbance to potentially other users obtaining corrections from the station. The evaluation is performed by analyzing the state of the \gls{rtk} baseline and the error under different adversarial conditions and receiver configurations.

\section{System and Adversary model}
Two \gls{gnss} receivers communicate over a \gls{ntrip} interface \cite{ntrip_esa}. The communication link is secured and the adversary cannot change data in transit, impersonate the reference (source of the stream), or in any way tamper with the \gls{ntrip} provided correction stream. The adversary can cause a Denial of Service (DoS), effectively making the rover unable to connect to the station but this is beyond the scope of this work as we assume the rover can at any moment connect, disconnect, and receive corrections from any available station. Additionally, we assume that the base station is honest and, unless adversarial manipulation is present, it provides legitimate, trustworthy corrections to any connected rover.

As the \gls{gnss} signal structure is known, the attacker can craft signals for any constellation or frequency (not cryptographically protected) that match the legitimate signals (modulation, frequency allocation, and data content). In addition, the adversary can generate signals so that the resulting \gls{pnt} solution at the station matches the attacker's objective. Given that stations are mounted at precise locations, the adversary has full knowledge of the type of receiver, antenna, and position of the phase center of the antenna to centimeter-level accuracy. With this knowledge, the attacker can control the \gls{gnss} receiver via spoofing or meaconing or cause denial of service by jamming.  

\section{Experiment Setup}
\label{sec:setup} 

\begin{figure}[h!]
    \centering
    \includegraphics[width=\linewidth]{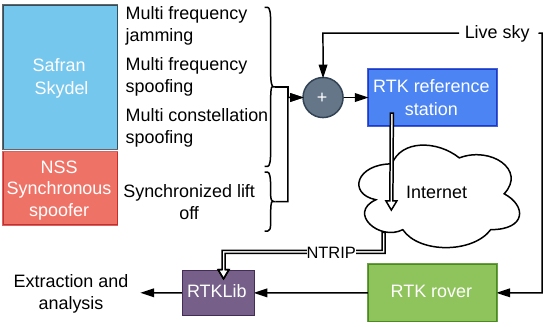}
    \caption{RTK test bed: implementation of station, rover and attacker.}
    \label{fig:rtkconfig-setup}
\end{figure}

The system setup considers two u-Blox ZED-F9P multi-frequency, high-precision \gls{gnss} receivers, each connected to a platform capable of providing connectivity and computation. One device is configured to broadcast \gls{rtk} corrections over a secure channel using a standard \gls{ntrip} provider to all connected clients. In \cref{fig:rtkconfig-setup}, this device is defined as \textit{station}. For the purpose of the demonstration in this work, it is not important how the rover and the reference station exchange information in a secure, authenticated way (e.g., this can be implemented with secure network transport). The rover receiver provides raw \gls{gnss} measurement data to an implementation of RTKLib (open source, at \cite{rtklibexplorer}) that processes the \gls{rtk} solution based on the correction stream obtained from the \gls{ntrip} server.

The adversary is implemented in two ways. First, we use a custom-made GPS L1 spoofer capable of code-phase-aligned and time-frame-aligned constellation coherent spoofing. This allows controlling the pseudorange of each satellite, by extending or shrinking it in a coordinated fashion and changing the time offset of the \gls{gnss} victim receiver. Additionally, this forces the station \gls{gnss} to produce fake pseudorange corrections at the \gls{ntrip} server. If the capture is successful, the adversary obtains full control of the \gls{rtk} station \gls{gnss} receiver.

Second, we use Safran Skydel to generate a set of different scenarios, including jamming with different signals and spoofing. Multiple constellations and frequency bands can be simulated and spoofed at the same time, showing that even multi-constellation and multi-frequency \gls{rtk} stations can be manipulated to produce valid but degraded correction streams. 

The experiments are conducted without radiating power in the \gls{gnss} frequency bands, following the local interference avoidance regulations - all tests are conducted either in a shielded environment or via cable. 

%%=============================================
\section{Evaluation and conclusions}
\label{sec:eval}

\begin{figure}[h!]
    \centering
    \includegraphics[width=\linewidth]{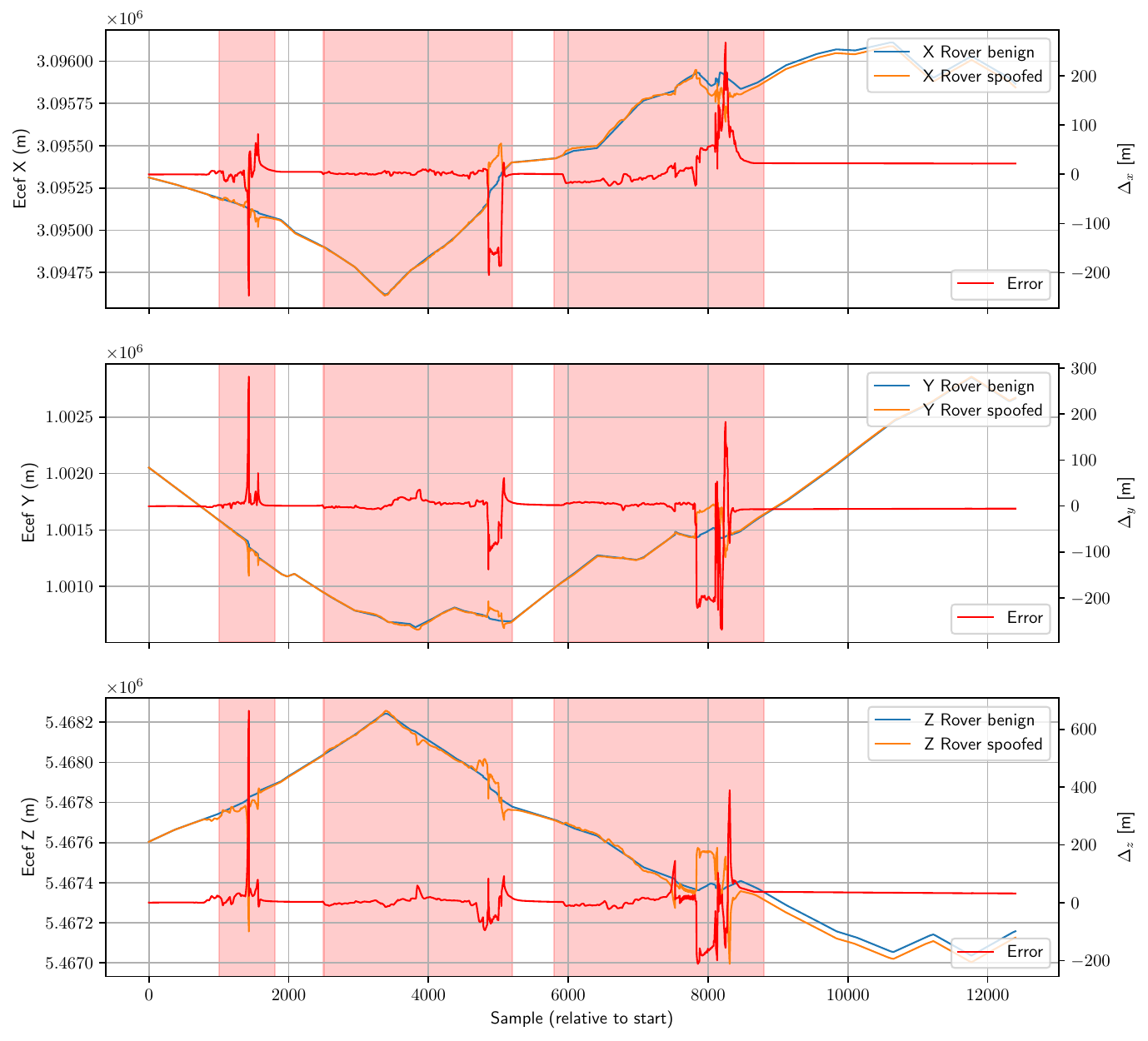}
    \caption{RTK degradation at the rover during synchronous multi-constellation spoofing.}
    \label{fig:rtk-rover-synchro}
\end{figure}

We evaluate three scenarios: synchronous spoofing, asynchronous spoofing and jamming of the reference station. The most interesting and advanced case is shown in \cref{fig:rtk-rover-synchro}. During synchronous multi-constellation spoofing with overpower, the reference station is captured during the multiple attempts (marked in red in \cref{fig:rtk-rover-synchro}). During the attack, the degradation is severe with a 3D-RMS error of more than \SI{50}{\meter} and significantly degraded altitude. The rover reaches convergence quickly after the spoofing action stops and rapidly recovers from the attack. Nevertheless, neither the GNSS receiver nor the RTKLib implementation seems to be aware of the spoofed reference station and instead of rejecting meaningless corrections, tries to reach convergence degrading the RTK solution quality which goes from full RTK fix (where the carrier phase information is fully resolved) to Differential GNSS (considering only double differences on the pseudoranges). Instead, the receiver should reject any correction that does not improve the accuracy achievable in stand-alone positioning mode. Additional investigations in these regards are ongoing.

Further results and test cases are shown in the graphical poster, analyzing the convergence ratio, other types of attacks and potentially a proposal for a countermeasure aiming at mitigating misbehaving reference stations.

\begin{acks}
    This work was supported by the Swedish Foundation for Strategic Research (SSF) SURPRISE project, the KAW Academy Fellow Trustworthy IoT project, and the Safran Minerva program. 
\end{acks}

\bibliographystyle{ACM-Reference-Format}
\bibliography{gnss-update}

\end{document}

%% file: acronyms.tex
\newacronym{gnss}{GNSS}{Global Navigation Satellite Systems}
\newacronym{gps}{GPS}{Global Positioning System}
\newacronym{nmea}{NMEA}{National Marine Electronics Association}
\newacronym{nasa}{NASA}{National Aeronautics and Space Administration}
\newacronym{agps}{A-GPS}{Assisted/Augmented GPS}
\newacronym{scer}{SCER}{Security Code Estimation and Replay}
\newacronym{nma}{OS-NMA}{Open Service Navigation Message Authentication}
\newacronym{sis}{SIS}{Signal In Space}
\newacronym{raim}{RAIM}{Receiver autonomous integrity monitoring}
\newacronym{pnt}{PNT}{Position-Navigation-Time}
\newacronym{rtk}{RTK}{Real-time Kinematics}
\newacronym{ntrip}{NTRIP}{Networked Transport of RTCM via Internet Protocol}
\newacronym{sdr}{SDR}{Software Defined Radio}
\newacronym{sdrs}{SDR}{Software Defined Radios}
\newacronym{ots}{OTS}{Off-the-Shelf}
\newacronym{rt}{RT}{Real-Time}
\newacronym{ide}{IDE}{Integrated Development Environment}
\newacronym{fifo}{FIFO}{First-In-First-Out}
\newacronym{sma}{SMA}{SubMiniature Version A}
\newacronym{rf}{RF}{Radio Frequency}
\newacronym{uav}{UAV}{Unmanned Aerial Vehicle}

%% file: Poster__testing_network_based_RTK_for_GNSS_robustenss_and_security.bbl

\begin{thebibliography}{8}

%%% the Web address.

\ifx \showCODEN    \undefined \def \showCODEN     #1{\unskip}     \fi
\ifx \showDOI      \undefined \def \showDOI       #1{#1}\fi
\ifx \showISBNx    \undefined \def \showISBNx     #1{\unskip}     \fi
\ifx \showISBNxiii \undefined \def \showISBNxiii  #1{\unskip}     \fi
\ifx \showISSN     \undefined \def \showISSN      #1{\unskip}     \fi
\ifx \showLCCN     \undefined \def \showLCCN      #1{\unskip}     \fi
\ifx \shownote     \undefined \def \shownote      #1{#1}          \fi
\ifx \showarticletitle \undefined \def \showarticletitle #1{#1}   \fi
\ifx \showURL      \undefined \def \showURL       {\relax}        \fi
% invisible to TeX
\providecommand\bibfield[2]{#2}
\providecommand\bibinfo[2]{#2}
\providecommand\natexlab[1]{#1}
\providecommand\showeprint[2][]{arXiv:#2}

\bibitem[\protect\citeauthoryear{(ESA)}{(ESA)}{2021}]%
        {ntrip_esa}
\bibfield{author}{\bibinfo{person}{European Space~Agency (ESA)}.}
  \bibinfo{year}{2021}\natexlab{}.
\newblock \bibinfo{booktitle}{\emph{Networked Transport of RTCM via Internet
  Protocol (Ntrip)}}.
\newblock
\urldef\tempurl%
\url{https://gssc.esa.int/wp-content/uploads/2018/07/NtripDocumentation.pdf}
\showURL{%
Retrieved March 11, 2024 from \tempurl}


\bibitem[\protect\citeauthoryear{GPSPatron}{GPSPatron}{2021}]%
        {gpspatron}
\bibfield{author}{\bibinfo{person}{GPSPatron}.}
  \bibinfo{year}{2021}\natexlab{}.
\newblock \bibinfo{booktitle}{\emph{How Non-Coherent Spoofing Affects GNSS Base
  Stations}}.
\newblock
\urldef\tempurl%
\url{https://gpspatron.com/how-non-coherent-spoofing-affects-gnss-base-stations/}
\showURL{%
Retrieved March 11, 2024 from \tempurl}


\bibitem[\protect\citeauthoryear{Humphreys, Ledvina, Psiaki, O'Hanlon, and
  Kintner}{Humphreys et~al\mbox{.}}{2008}]%
        {humphreysAssessingSpoofingThreat2008}
\bibfield{author}{\bibinfo{person}{T.~E. Humphreys}, \bibinfo{person}{B.~M.
  Ledvina}, \bibinfo{person}{M.~L. Psiaki}, \bibinfo{person}{B.~W. O'Hanlon},
  {and} \bibinfo{person}{P.~M. Kintner}.} \bibinfo{year}{2008}\natexlab{}.
\newblock \showarticletitle{{Assessing the spoofing threat: Development of a
  portable GPS civilian spoofer}}. In \bibinfo{booktitle}{\emph{ION GNSS}}
  (Savannah, GA, USA).
\newblock


\bibitem[\protect\citeauthoryear{Lenhart, Spanghero, and
  Papadimitratos}{Lenhart et~al\mbox{.}}{2022}]%
        {LenhartSP:C:2022}
\bibfield{author}{\bibinfo{person}{M. Lenhart}, \bibinfo{person}{M Spanghero},
  {and} \bibinfo{person}{P. Papadimitratos}.} \bibinfo{year}{2022}\natexlab{}.
\newblock \showarticletitle{{Distributed and Mobile Message Level
  Relaying/Replaying of GNSS Signals}}. In
  \bibinfo{booktitle}{\emph{International Technical Meeting of The Institute of
  Navigation (ITM)}}. \bibinfo{address}{Long Beach, CA, USA}.
\newblock


\bibitem[\protect\citeauthoryear{Motallebighomi, Sathaye, Singh, and
  Ranganathan}{Motallebighomi et~al\mbox{.}}{2023}]%
        {Motallebighomi2023}
\bibfield{author}{\bibinfo{person}{M. Motallebighomi}, \bibinfo{person}{H.
  Sathaye}, \bibinfo{person}{M. Singh}, {and} \bibinfo{person}{A.
  Ranganathan}.} \bibinfo{year}{2023}\natexlab{}.
\newblock \showarticletitle{Location-independent GNSS Relay Attacks: A Lazy
  Attacker's Guide to Bypassing Navigation Message Authentication}. In
  \bibinfo{booktitle}{\emph{ACM Conference on Security and Privacy in Wireless
  and Mobile Networks}}. \bibinfo{address}{Guildford, UK}.
\newblock


\bibitem[\protect\citeauthoryear{Papadimitratos and Jovanovic}{Papadimitratos
  and Jovanovic}{2008}]%
        {papadimitratosProtectionFundamentalVulnerability2008}
\bibfield{author}{\bibinfo{person}{P. Papadimitratos} {and} \bibinfo{person}{A.
  Jovanovic}.} \bibinfo{year}{2008}\natexlab{}.
\newblock \showarticletitle{{Protection and Fundamental Vulnerability of
  GNSS}}. In \bibinfo{booktitle}{\emph{IEEE IWSSC}}.
  \bibinfo{address}{Toulouse, France}.
\newblock


\bibitem[\protect\citeauthoryear{RTKLibExplorer}{RTKLibExplorer}{2021}]%
        {rtklibexplorer}
\bibfield{author}{\bibinfo{person}{RTKLibExplorer}.}
  \bibinfo{year}{2021}\natexlab{}.
\newblock \bibinfo{booktitle}{\emph{A version of RTKLIB optimized for single
  and dual frequency low cost GPS receivers, especially u-blox receivers.}}
\newblock
\urldef\tempurl%
\url{https://github.com/rtklibexplorer/RTKLIB}
\showURL{%
Retrieved March 11, 2024 from \tempurl}


\bibitem[\protect\citeauthoryear{van Tol}{van Tol}{2020}]%
        {Pepjin:2020}
\bibfield{author}{\bibinfo{person}{Pepijn van Tol}.}
  \bibinfo{year}{2020}\natexlab{}.
\newblock \emph{\bibinfo{title}{RTK-GNSS augmentation data spoofing}}.
\newblock \bibinfo{thesistype}{Master's\ thesis}. \bibinfo{address}{Delft
  University of Technology, The Netherlands}.
\newblock


\end{thebibliography}
